\documentclass[twocolumn,pra,superscriptaddress,showpacs]{revtex4}
\usepackage{times}
\usepackage{color}
\usepackage{amsmath}
\usepackage{graphicx}
\usepackage{amssymb}
\usepackage{esint}

\begin{document}
\title{Polariton states in circuit QED for electromagnetically induced transparency}
\author{Xiu Gu}
\affiliation{Institute of Microelectronics, Tsinghua University, Beijing 100084, China}
\affiliation{CEMS, RIKEN, Saitama 351-0198, Japan}
\author{Sai-nan Huai}
\affiliation{Institute of Microelectronics, Tsinghua University, Beijing 100084, China}
\author{Franco Nori}
\affiliation{CEMS, RIKEN, Saitama 351-0198, Japan}
\affiliation{Department of Physics, The University of Michigan, Ann Arbor, Michigan
48109-1040, USA}
\author{Yu-xi Liu}
\email{yuxiliu@mail.tsinghua.edu.cn}
\affiliation{Institute of Microelectronics, Tsinghua University, Beijing 100084, China}
\affiliation{Tsinghua National Laboratory for Information Science and Technology
(TNList), Beijing 100084, China}
\affiliation{CEMS, RIKEN, Saitama 351-0198, Japan}
\date{\today}

\begin{abstract}
Electromagnetically induced transparency (EIT) has been extensively studied in various systems. However, it is not easy to observe in superconducting quantum circuits (SQCs), because the Rabi frequency of the strong controlling field corresponding to EIT is limited by the decay rates of the SQCs. Here, we show that EIT can be achieved by engineering decay rates in a superconducting circuit QED system through a classical driving field on the qubit. Without such a driving field, the superconducting qubit and the cavity field are approximately decoupled in the large detuning regime, and thus the eigenstates of the system are approximately product states of the cavity field and qubit states. However, the driving field can strongly mix these product states and so-called polariton states can be formed.  The weights of the states for the qubit and cavity field in the polariton states can be tuned by the driving field, and thus the decay rates of the polariton states can be changed. We choose the three lowest-energy polariton states with $\Lambda$-type transitions in such a driven circuit QED system, and demonstrate how EIT and ATS can be realized in this compound system. We believe that this study will be helpful for EIT experiments using SQCs. 
\end{abstract}

\maketitle
\section{Introduction}

Since electromagnetically induced transparency (EIT)
was proposed~\cite{Harris1990}, it has been extensively explored in various contexts~\cite{Harris1997,Marangos1998,Lukin2003,Fleischhauer2005} using three-level systems. The main feature of EIT is that the absorption of of a weak probe field in a medium is
reduced because of the presence of a strong control field. EIT can be used to control the propagation of the weak field through the medium. It can also be used to greatly enhance the nonlinear
susceptibility in the induced transparency region, and thus to generate a strong
photon-photon Kerr interaction. Strong photon-photon Kerr interactions have been studied to realize quantum logic operations such as controlled phase gates~\cite{Turchette1995}, quantum Fredkin gates~\cite{Milburn1989} and conditional phase switches~\cite{Resch2002} for photon-based quantum information processing. Moreover, Kerr interactions can be employed to realize quantum nondemolition detection of photons~\cite{Imoto1985}.

Recent studies show that superconducting quantum circuits (SQC) are one of the best candidates for quantum information
processing~\cite{You2005,Clarke2008,Schoelkopf2008,You2011,Devoret2013}. Meanwhile, these artificial atoms~\cite{You2005,Clarke2008,Buluta2010,You2011,Xiang2013} have been employed to study quantum optics and atomic physics in the microwave domain. For example, several studies~\cite{Zhou2002,Amin2003,Kis2004,Yang2004,Kelly2010} have explored population trapping and dark states in three-level SQCs.  EIT
was also theoretically studied for probing the decoherence of a superconducting flux qubit~\cite{Murali2004,Dutton2006} via a third auxiliary state.  How to realize EIT using SQCs has also been theoretically studied using several different setups~\cite{Ian2010,Joo2010,SunHuiChen2014}. Experimentalists showed Autler-Townes splitting (ATS)~\cite{Autler1955} using various three-level SQCs~\cite{Baur2009,Sillanpaa2009,Abdumalikov2010,Kelly2010,Li2011a,Li2012,Hoi2013,Novikov2013,Hoi2013a,Suri2013}. However, to our knowledge, up to now, there is no experiment on EIT using SQCs. The main obstacle is that the decay rates of the three-level system and the strength of the Rabi frequency corresponding to the controlling field cannot satisfy the condition for realizing EIT~\cite{Abi-Salloum2010a,Anisimov2011,SunHuiChen2014}.

It is well known that EIT~\cite{Harris1990} is mainly caused by Fano interference~\cite{Fano1961}, while ATS~\cite{Autler1955} is due to the driving-field-induced shift of the transition frequency. Although the mechanisms of EIT and ATS are very different, they are not easy to discern from experimental observations, since both of them exhibit a dip in the absorption spectrum of the weak probe field. Theoretically, there is a threshold value~\cite{SunHuiChen2014,Abi-Salloum2010a,Anisimov2011} to distinguish EIT from ATS. This threshold value is determined by two decay rates of the three-level system~\cite{SunHuiChen2014,Abi-Salloum2010a,Anisimov2011}. When the strength of the Rabi frequency of the strong controlling field is smaller than this value, EIT occurs, otherwise, it is ATS. Experimentally, the data should be analyzed by virtue of the Akaike information criterion~\cite{Anisimov2011}.  The transition from EIT to ATS has been experimentally demonstrated in coupled whispering-gallery-mode optical resonators~\cite{Peng2014}.

Here, \textit{instead of three-level} superconducting quantum circuits~~\cite{Murali2004,Dutton2006,Ian2010,Joo2010,SunHuiChen2014},
we study EIT and the transition from EIT to ATS using \textit{a driven two-level} circuit QED system~\cite{Schoelkopf2008}, where a superconducting qubit is coupled to a single-mode cavity field and driven by a classical field. The three-level system used to study EIT is constructed by the three lowest-energy mixed polariton states, formed by the driving field and the states of the circuit QED system. The polariton states are hybridizations of microwave photon and qubit states. Thus, the decays of the polariton states are determined by the decays of both the cavity field and the superconducting qubit.

The eigenstates of the circuit QED system are mixed states of the cavity field and the superconducting qubit. These states can be approximately reduced to product states of the uncoupled cavity field and qubit states when the frequencies of the cavity field and the superconducting qubit are largely detuned.  In this case, the qubit acquires a small frequency shift due to the microwave field and the Purcell~\cite{Purcell1946} enhanced spontaneous decay rate~\cite{Houck2007} is obtained. Therefore, when the detuning between the cavity field and the qubit is changed from zero to a finite value, the decay rates of the eigenstates of the circuit QED system can also be changed. However, such tunable decay is not easy to be realized when the sample is fabricated. Thus, we introduce a classical driving field to further mix the eigenstates of the circuit QED. We call these newly mixed states as polariton states.  The weights of the photon and qubit states in polariton states can be tuned by the driving field, and thus the decay of the polariton states can be controlled. Such polariton states were studied in order to implement an impedance-matched $\Lambda$ system~\cite{Koshino2013a,Koshino2013,Inomata2014}, where the two decay rates from the top energy level to the two lowest energy levels are identical. However, in our study, we need to engineer different decay rates of the three-level $\Lambda$ system so that the condition to realize EIT and ATS can be satisfied. In contrast to the study~\cite{Ian2010}, an extra driving field on the qubit is introduced to modify the decay rates of the system studied.

The paper is organized as follows. In Sec.~II, we describe a model Hamiltonian and discuss how the transition frequencies
can be tuned by the driving field. In Sec.III, we study the selection rules and how to control decay rates of
polariton states by changing driving field. In Sec.~IV, we study a three-level polariton system to implement EIT and ATS, and the
threshold value to discern EIT and ATS is given. Numerical simulations with possible experimental parameters are presented. Finally, further discussions and a summary are presented in Sec.~V.

\begin{figure}[b]
\includegraphics[scale=0.38]{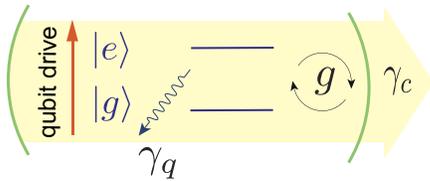}
\caption{ (color online). A driven qubit coupled to a resonator mode. Here, $\{|g\rangle$,$|e\rangle\}$ are the ground and excited states of the qubit,
$g$ is the coupling strength between the qubit and the cavity field, $\gamma_q$ ($\gamma_c$) is the decay rate of the qubit
(cavity field). 
}\label{fig1}
\end{figure}
\begin{figure}[b]
\includegraphics[width=8cm]{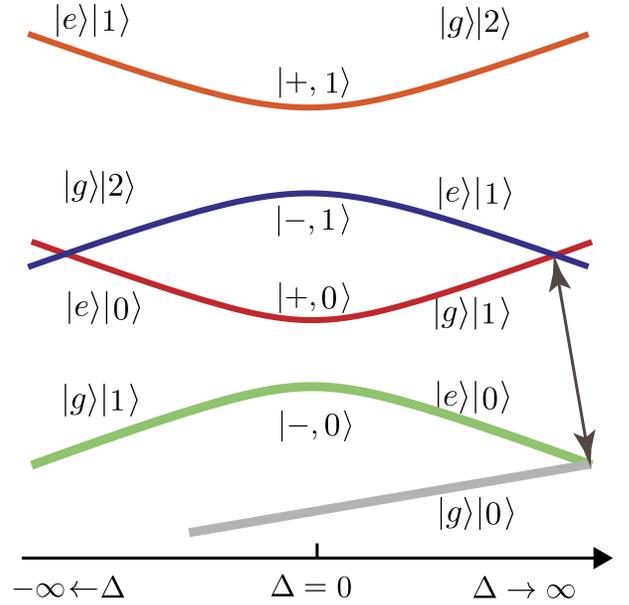}
\caption{ (color online). A schematic energy diagram for the Jaynes-Cummings model versus the detuning between the qubit and cavity field in the rotating frame. The eigenstates are denoted by $|\pm,n\rangle$. In the large-detuning regime, the coupled states $|\pm,n\rangle$ are approaching the bare qubit and photon states. Here, $\Delta=\tilde\omega_r-\tilde\omega_q$. The nesting regime, defined by $\omega_{|g,0\rangle} \le\omega_{|e,0\rangle} <\omega_{|e,1\rangle}\le\omega_{|g,1\rangle}$, as in Refs.~\cite{Inomata2014,Koshino2013a,Koshino2013}, is between the two points linked by a line with two arrows. The frequency of the driving field $\omega_d$ sets the boundary of the nesting regime. The lower limit is at $\omega_{|e,1\rangle} = \omega_{|g,1\rangle}$, where $\omega_d=\omega_q-3\chi$. The upper limit is at $\omega_{|g,0\rangle} = \omega_{|e,0\rangle}$, where $\omega_d=\omega_q-\chi$.  Here $\chi=g^2/\Delta$, with $g$ the coupling strength between the qubit and cavity field. We note that when the zero-point fluctuation of the cavity field is taken into account, then the energy of the ground state $|g\rangle|0\rangle$ is $\Delta/2$. }\label{fig2}
\end{figure}

\section{Theoretical model and polariton states}

In this section, we derive polarition states from the model Hamiltonian and then discuss how decay rates of the polariton states can be adjusted by an externally-applied classical field.

\subsection{Hamiltonian}

As schematically shown in Fig.~\ref{fig1}, we study a superconducting two-level system (a qubit system) which is coupled to a quantized single-mode microwave field and also driven by a classical microwave field. For concreteness, we assume that such qubit system is a three-Josephson-junction flux qubit circuit.  The interaction between the qubit and the single-mode cavity field is described by the well-known Jaynes-Cummings model~\cite{JaynesCummings,Shore1993}.  Thus, the model Hamiltonian of the driven circuit QED system can be written as
\begin{eqnarray}\label{eq:1}
H_{S} &=&\frac{\hbar }{2}\omega _{q}\sigma _{z}+\hbar\omega_{r} \left(a^{\dag }a+\frac{1}{2}\right)+\hbar g\left(a^{\dag }\sigma _{-}+a\sigma _{+}\right)\\ \nonumber
&+&\hbar \left[ \Omega\sigma _{-}\exp\left(i\omega_{d} t\right)+\Omega ^{\ast }\sigma _{+}\exp\left(-i\omega_{d} t\right)\right].
\end{eqnarray}
The first line of Eq.~(\ref{eq:1}) is the Jaynes-Cummings Hamiltonian, which describes the interaction between the qubit system and the single-mode cavity field with coupling strength $g$. Here $\omega _{q}$ and $\omega _{r}$ denote the frequencies of the qubit and single-mode cavity field, respectively. Also, $a$ is the annihilation operator of the cavity field and $\sigma_-$ is the  ladder operator of the qubit. The second line of Eq.~(\ref{eq:1}) describes the interaction between the qubit and the classical driving field. The parameter $\Omega$ represents the interaction strength or Rabi frequency between the qubit and the classical field with frequency $\omega_{d}$. Without loss of generality, hereafter we assume that $\Omega$ is a real number.

To remove the time-dependent factors, we transform the Hamiltonian $H_{S}$, given in Eq.~(\ref{eq:1}), into a rotating reference frame by the unitary transformation
\begin{equation}
U=\exp[-i\omega_{d}\left(\sigma_{z}/2+a^{\dagger}a\right)t]
\end{equation}
so that we can obtain the following effective Hamiltonian
\begin{eqnarray}\label{eq:Hr}
\tilde{H}_{S} &=&\frac{\hbar }{2}\tilde{\omega}_{q}\sigma _{z}+\hbar\tilde{\omega}_{r} \left(a^{\dag }a+\frac{1}{2}\right)+\hbar g\left(a^{\dag }\sigma _{-}+a\sigma _{+}\right) \nonumber \\
&+&\hbar \lbrack \Omega\sigma _{-}+\Omega\sigma _{+}]
\end{eqnarray}
with the detunings $\tilde{\omega}_{q}=\omega _{q}-\omega _{d}$, $\tilde{\omega}_{r}=\omega _{r}-\omega _{d}$, and $\Delta=\tilde\omega_{r}-\tilde\omega_{q}$.

\subsection{Eigenvalues and eigenstates for $\Omega=0$}
For completeness, we first briefly discuss the eigenstates and eigenvaules when the classical driving field is not applied to the qubit. In this case, $\Omega=0$, the eigenstates of Eq.~(\ref{eq:Hr}), for the Jaynes-Cummings Hamiltonian of the qubit and the single-mode cavity field, are
\begin{eqnarray}
|+,n\rangle&=&\cos\frac{\theta_{n}}{2}|e,n\rangle+\sin\frac{\theta_{n}}{2}|g,n+1\rangle,\label{eq:4}\\
|-,n\rangle&=&-\sin\frac{\theta_{n}}{2}|e,n\rangle+\cos\frac{\theta_{n}}{2}|g,n+1\rangle,\label{eq:5}
\end{eqnarray}
which mix the qubit states  with the states  of a single-mode cavity field. Here, $\tan\theta_{n}=-2g\sqrt{n+1}/\Delta$.
We note
that $|e,n\rangle\equiv |e\rangle|n\rangle$ ($|g,n\rangle\equiv|g\rangle|n\rangle$) denote that the qubit is in the excited $|e\rangle$ (ground $|g\rangle$) state and the single-mode cavity field is in the state $|n\rangle$.
The states expressed in Eqs.~(\ref{eq:4}) and (\ref{eq:5}) are usually called dressed states. Note that we do not distinguish the dressed states in the rotating reference frame from those in original laboratory frame.
The eigenvalues corresponding to Eqs.~(\ref{eq:4}) and (\ref{eq:5}) are
\begin{equation}
E_{\pm,n}=\hbar\tilde{\omega}_{r}\left(n+\frac{1}{2}\right)\pm \frac{\hbar}{2}\sqrt{\Delta^2+4g^2\left(n+1\right)}
\label{eq:6}
\end{equation}

\begin{figure}[b]
\includegraphics[scale=1]{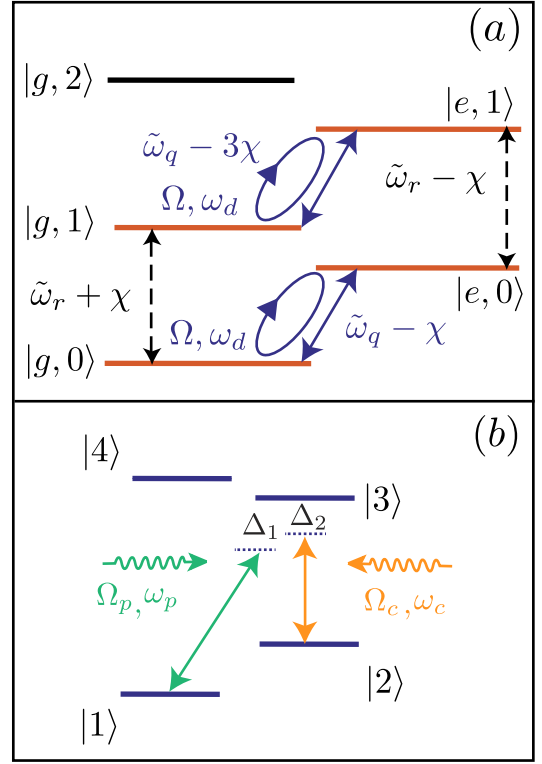}
\caption{ (color online). (a)  The dressed states of the Jaynes-Cummings mode for large detuning is mixed further by a driving field applied to the qubit, which only connects $|g,n\rangle$ and $|e,n\rangle$. In the un-nesting regime, without the driving field, the lowest four levels are approximately $|g,0\rangle,|e,0\rangle,|g,1\rangle,|e,1\rangle$, which are mixed when the driving field is applied to the qubit. (b) Four polariton states, i.e., energy levels $|i\rangle$ with ($i=1,2,3,4$) expressed in Eqs.~(\ref{eq:11}-\ref{eq:12}),  and Eqs.(\ref{eq:14}~\ref{eq:15}), in which we choose the three lowest energy levels to study EIT and ATS in Sec.~IV. } \label{fig3}
\end{figure}

From Eqs.~(\ref{eq:4}) and (\ref{eq:5}), it is clear that the eigenenergies of the Jaynes-Cummings model change with the detuning $\Delta$ between the qubit and the single-mode cavity field.  When the detuning is very large, the dressed states are approaching either bare qubit states or the states of the single-mode cavity field. That is, they are almost decoupled from each other. The eigenenergies in Eq.~(\ref{eq:6}) are shown  as a function of the detuning $\Delta$ in Fig.~\ref{fig2}, which clearly shows that the qubit and the cavity field are decoupled from each other when $\Delta$ becomes very large. Figure~\ref{fig2} also shows that there are some degeneracy points when $\Delta$ takes a particular value, e.g., $E_{g,0}=E_{-,0}$ and $E_{+,0}=E_{-,1}$, which will be further discussed in the following in the large-detuning case.

In the large-detuning case, i.e., $g\ll|\Delta|$, Eqs.~(\ref{eq:4}) and (\ref{eq:5}) can be approximately written as
\begin{eqnarray}
|+,n\rangle & \approx &|g,n+1\rangle-\frac{g}{\Delta}\sqrt{n+1}|e,n\rangle,\label{eq:7}\\
|-,n\rangle & \approx &|e,n\rangle+\frac{g}{\Delta}\sqrt{n+1}|g,n+1\rangle,\label{eq:8}
\end{eqnarray}
where we assume $\Delta>0$ for convenience in the following discussions.
We now focus on the five lowest eigenstates of the Jaynes-Cummings model, i.e., the ground state $|g,0\rangle$, and the four dressed states $|\pm, 0\rangle$ and $|\pm, 1\rangle$.  To simplify the analysis, we first omit the first order of $g/\Delta$. In this case,  the four dressed states can be approximately written as $|-,0\rangle \approx |e,0\rangle$, $|+,0\rangle\approx |g,1\rangle$, $|-,1\rangle\approx |e,1\rangle$, $|+,1\rangle\approx  |g,2\rangle$, which correspond to the eigenfrequencies $E_{\pm,n}/\hbar$ given by
\begin{eqnarray}
\omega _{|g,n\rangle }&\approx& n\left(\tilde{\omega}_{r}+\chi\right)+\frac{\Delta}{2},\label{eq:9}\\
\omega _{|e,n\rangle }&\approx& \tilde{\omega}_q-\chi+n\left(\tilde{\omega}_{r}-\chi\right) +\frac{\Delta}{2},\label{eq:10}
\end{eqnarray}
where $\chi=g^{2}/\Delta$ is the dispersive frequency shift.
From Eqs.~(\ref{eq:9}) and (\ref{eq:10}), we can approximately obtain $\omega_{|e,0\rangle}=\omega_{|g,0\rangle}$ at $\tilde{\omega}_q=\chi$, and $\omega_{|e,1\rangle}=\omega_{|g,1\rangle}$ at $\tilde{\omega}_q=3\chi$. Thus, to operate in the so-called nesting regime~\cite{Inomata2014,Koshino2013a,Koshino2013}, where $\omega_{|g,0\rangle} <\omega_{|e,0\rangle} <\omega_{|e,1\rangle}<\omega_{|g,1\rangle}$, that is, $E_{g,0}<E_{-,0}<E_{-,1}<E_{+,0}$, the frequency $\omega_d$ of the driving field must satisfy the condition $\omega_q-3\chi<\omega_d<\omega_q-\chi$.

\subsection{Eigenvalues and eigenstates for $\Omega\neq 0$}
When a classical driving field is applied to the qubit, i.e., $\Omega\neq 0$,  then it will induce transitions between different states of $|\pm ,n\rangle$. Thus, the classical driving field lifts the degeneracies and strongly mixes the states $|\pm,n\rangle$ with the states $|\pm,n+1\rangle$ . That is, the dressed states in Eqs.~(\ref{eq:4}) and (\ref{eq:5}) are mixed again by the classical field. We call these new states as polariton states, because they inherit both atomic and photonic properties. Below, we will mainly focus on the large- detuning regime.

As schematically shown in Fig.~\ref{fig3} for the large detuning case, where the first-order term in the parameter $g/(\omega_{r}-\omega_{q})$ for the dressed states are omitted, the four lowest states discussed above Eq.~(\ref{eq:9}) are mixed by the classical field, i.e., the qubit is doubly
 dressed by a single-mode cavity field and a classical driving field.
 In the nesting regime, as shown in Refs.~\cite{Inomata2014,Koshino2013a,Koshino2013}, a weak driving field, applied to the qubit, can drastically change the ratio of the contributions from $|g,n\rangle$ and $|e,n\rangle$ to the final polariton states. This is in contrast with the un-nesting case, where the external qubit drive has no appreciable effect on the system as the following analysis shows.

The classical qubit field only induces transitions between the states $|g,n\rangle$
and $|e,n\rangle$. Thus, it can only mix the state $|g,0\rangle$ with $|\pm,0\rangle$, or mix states $|\pm,n\rangle$ with states $|\pm, n+1\rangle$. In the large-detuning case, the states $|g,0\rangle$ and $|-,0\rangle\approx |e,0\rangle$ are separated by the energy level spacing
$\omega_q-\chi$, when the driving field is not applied. Thus, in the lower boundary of the nesting regime, when the frequency of the driving field satisfies the condition
$\omega_d=\omega_q-\chi$, the driving field induce transitions between the states $|-,0\rangle\approx |e,0\rangle$ and $|g,0\rangle$ and strongly mix these two states. These mixed states form new doubly-dressed eigenstates, the so-called polariton states,
 \begin{eqnarray}
|1\rangle&=&-\sin \frac{\theta _{l}}{2}|e,0\rangle+\cos \frac{\theta _{l}}{2}|g,0\rangle,\label{eq:11}\\
|2\rangle&=&\cos \frac{\theta _{l}}{2}|e,0\rangle+\sin \frac{\theta _{l}}{2}|g,0\rangle.\label{eq:12}
\end{eqnarray}
Here $\tan\theta_l={2\Omega}/\left({\tilde{\omega}_q-\chi}\right)$. The transition frequency between the state $|1\rangle$ and the state $|2\rangle$ is given by
\begin{equation}\label{eq:13}
\omega_{21}=\sqrt{\left(\tilde\omega_{q}-\chi\right)^{2}+4\Omega^{2}},
\end{equation}
with $\omega_{ij}=\omega_i-\omega_j$.
Likewise,  the states $|+,0\rangle\approx|g,1\rangle$ and $|-,1\rangle\approx|e,1\rangle$, with original level spacing $\omega_q-3\chi$, can be mixed by the qubit driving field at the upper boundary of the nesting regime when
$\omega_d=\omega_q-3\chi$. Thus, these polariton states (i.e. eigenstates) can be given by
\begin{eqnarray}
|3\rangle&=&-\sin \frac{\theta _{u}}{2}|g,1\rangle+\cos \frac{\theta _{u}}{2}|e,1\rangle,\label{eq:14}\\
|4\rangle&=&\cos \frac{\theta _{u}}{2}|g,1\rangle+\sin \frac{\theta _{u}}{2}|e,1\rangle,\label{eq:15}
\end{eqnarray}
with $\tan\theta_u={2\Omega}/({-\tilde{\omega}_q+3\chi})$.
The energy splitting between $|4\rangle$ and $|3\rangle$ becomes
\begin{equation}\label{eq:16}
\omega_{43}= \sqrt{\left(\tilde\omega_{q}-3\chi\right)^{2}+4\Omega^{2}}.
\end{equation}
As schematically shown in Fig.~\ref{fig2}(b), below, we will choose $\{|1\rangle,|2\rangle,|3\rangle\}$ to form a three-level system. The transition frequency $\omega_{31}$  between the state $|3\rangle$ and the state $|1\rangle$ is given by
\begin{equation}\label{eq:17}
\omega_{31}=\tilde{\omega}_r-\frac{1}{2}\left(\omega_{43}+\omega_{21}\right).
\end{equation}
\begin{figure}[h]
	\includegraphics[width=8cm]{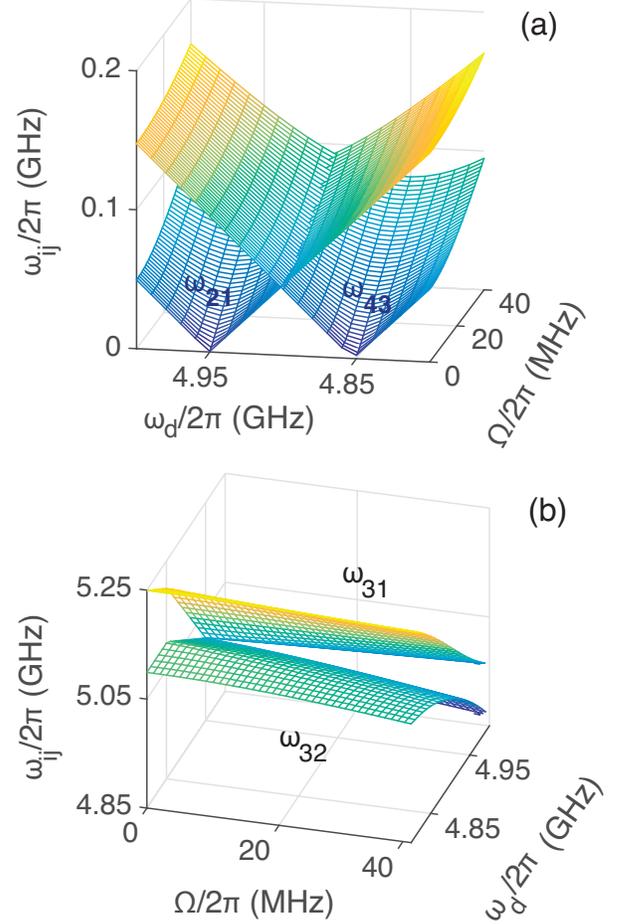}
	\caption{ (color online). Transition frequencies $\omega_{ij}$ versus both the frequency $\omega_d$ and the strength $\Omega$ of the driving field in the rotating reference frame. Here $\omega_{ij}=\omega_i-\omega_j$. The paremeters used here are $\omega_{q}/2\pi$=5 GHz, $\omega_{r}/2\pi$=10 GHz, $g/2\pi$=500 MHz, $\chi$=50 MHz. (a) Two V- shaped surfaces with the minimum values at $\omega_d=4.95$ GHz and $\omega_d=4.85$ GHz represent $\omega_{21}$ and $\omega_{43}$, respectively. (b) The upper surface represents $\omega_{31}$, the lower one represents $\omega_{32}$. }\label{fig4}
\end{figure}

In Fig.~\ref{fig4}, transition frequencies are plotted as functions of the qubit driving frequency $\omega_{d}$ and strength $\Omega$ in the rotating reference frame. Here we use the exact solution of Eq.~(\ref{eq:Hr}) in the numerical calculation. In other words, without large-dispersive approximation, each new eigenstate $|i\rangle$ is the superposition of five states: $|g,0\rangle$, $|e,0\rangle$, $|g,1\rangle$, $|e,1\rangle$, and $|g,2\rangle$.
Figure~\ref{fig4}(a) shows the trend of  $\omega_{21}$ and $\omega_{43}$, and they are consistent with Eqs.~(\ref{eq:13}) and ~(\ref{eq:16}). First we discuss the $\Omega=0$ case, in which the driving field is not applied. The lowest two energy levels $|1\rangle$ and  $|2\rangle$ are composed of $|g,0\rangle$ and $|e,0\rangle$. The degeneracy point, where $\omega_{21}=0$, is set by $\omega_{|e,0\rangle}=\omega_{|g,0\rangle}$, i.e. $\omega_d=\omega_q-\chi$, which is the upper boundary of the nesting regime. Likewise, the states $|4\rangle$ and $|3\rangle$ are superpositions of the states $|g,1\rangle$ and $|e,1\rangle$.
The degeneracy point is set by $\omega_{|e,1\rangle}=\omega_{|g,1\rangle}$, i.e. $\omega_d=\omega_q-3\chi$, which is the lower boundary of nesting regime. We note that $\omega_{43}=\omega_{21}$ in the middle of nesting regime when $\tilde\omega_q=2\chi$. It is clear that the frequency $\omega_d$ of the driving field determines the onset of the nesting regime. When the driving field is applied to the qubit, i.e., $\Omega\neq0$, these degeneracies are lifted. The larger strength $\Omega$ is, the larger $\omega_{21}$ and $\omega_{43}$ are.

We also show how the frequency and the strength of the driving field affect the transition frequencies $\omega_{31}$ and $\omega_{32}$ in Fig.~\ref{fig4}(b). It clearly shows that the upper surface for the transition frequency $\omega_{31}$ approaches that of $\omega_{32}$ when the states $|2\rangle$ and $|1\rangle$ are degenerate. When the driving strength is small, $\omega_{31}$ and $\omega_{21}$ are mainly determined by $\tilde\omega_r$ and $\chi$. However, $\chi$, which is determined by $g$ and $\Delta$  in the large-dispersive regime, can be enhanced by the presence of the higher excited states ~\cite{Houck2007,Inomata2014,Koshino2013a,Koshino2013}.
\section{Transition rules and tunable decay rates}

\subsection{Transition rules between polariton states}
Since the polariton states, formed by the cavity field, driving field, and the qubit, are mixed photon and qubit states, we can induce transitions between two of these new states by applying additional classical fields to either the qubit or the cavity field.  Hereafter we call these classical fields as the external fields, to avoid confusion with the classical driving field applied to the qubit with coupling strength $\Omega$.  The transition selection rule between these polariton states depends on the manner on how the external field is applied.  For example, the transitions between the state $|e,n\rangle$ and $|g,n\rangle$ can be induced when the external field is applied to the qubit. However, those transitions are forbidden when the external field is applied to the cavity field. In contrast, the transitions between the state $|l,n\rangle$  and $|l,n-1\rangle$, with $l=e$ or  $l=g$, can be induced by the external field applied to the cavity field. However, these transitions are forbidden if the external field is applied to the qubit. Therefore, the transition matrix elements between two polariton states can be tuned by varying the applied external field.

When the external field is applied to the qubit, the transition matrix elements are denoted by 
\begin{equation}
\text{Q}_{ij}={ \large |\langle i |\sigma_{-}| j\rangle|}.
 \end{equation} 
Similarly, when the external field is applied to the cavity field, the transition matrix elements are defined as 
\begin{equation}
\text{C}_{ij}={\large |\langle i| a |j\rangle |}.
\end{equation}
Here, $|i\rangle$ and $|j\rangle$ denote the new polariton states, e.g., the states in Eqs.~(\ref{eq:11}--\ref{eq:12}) and Eqs.~(\ref{eq:14}--\ref{eq:15}) in the large-detuning case. The transition elements between two of the states in Eqs.~(\ref{eq:11}--\ref{eq:12}) and (\ref{eq:14}), can be written as
\begin{eqnarray}\label{eq:Decay}
\text{C}_{32}& =& \left|\cos\large\left(\frac{\large\theta_{u}+\large\theta_{l}}{2}\right)\right|,\label{eq:18}\\
\text{C}_{31}& =& \left|\sin\large\left(\frac{\theta_{u}+\theta_{l}}{2}\right)\right|,\label{eq:19}\\
\text{Q}_{21}& = & \cos^{2}\left(\frac{\theta_{l}}{2}\right),\label{eq:20}\\
\text{Q}_{31} &= &\text{Q}_{32} = \text{C}_{21}= 0.\label{eq:21}
\end{eqnarray}

\begin{figure}[b]
	\includegraphics[scale=0.45]{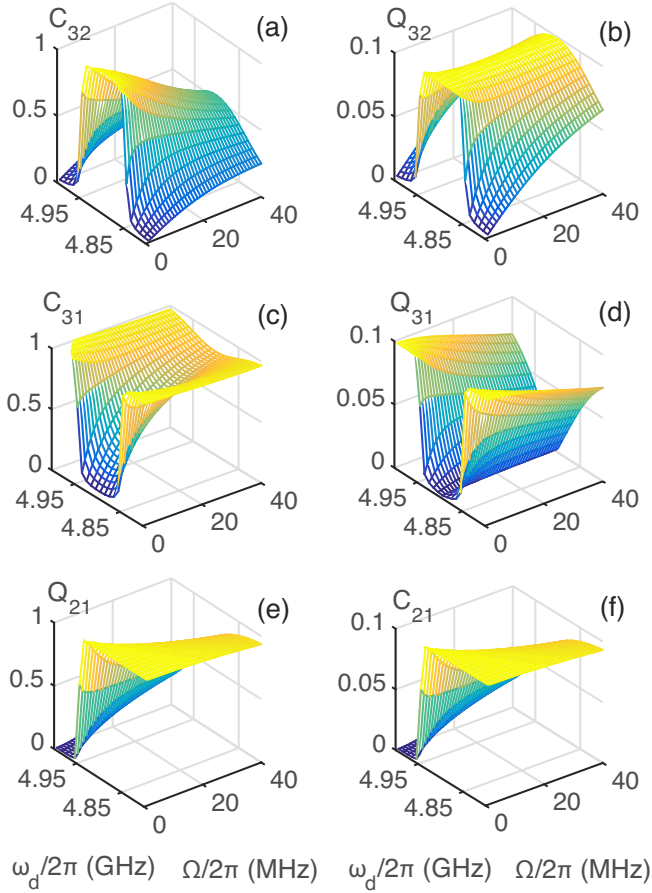}
	\caption{(color online). Moduli of the transition matrix elements between the state $|i\rangle$ and the state $|j\rangle$ versus both the strength $\Omega$ and the frequency $\omega_d$ of the driving field. $\text{Q}_{ij}$ denotes transition matrix elements between the states $|i\rangle$ and $|j\rangle$, induced by an external field applied to the qubit, while $\text{C}_{ij}$ is the one induced by an external field applied to the cavity field.
		The sharp change in $\text{Q}_{ij}$ and $\text{C}_{ij}$ occur at the boundary of the nesting regime; when $\Omega=0$ this is occurs when
		$\omega_q-3\chi<\omega_d<\omega_q-\chi$.  Note that the nonzero coupling in (b), (d), (e)  is limited
		by the first order of $g/\Delta$, which we omit in the theoretical analysis. The paremeters  are the
		same as in Fig.~\ref{fig4}.  Here, the states $|i\rangle$ for numerical calculations of the transition matrix elements are
		given by the exact eigenstates of Eq.~(\ref{eq:Hr}).}%
	\label{fig5}%
\end{figure}

Equations.~(\ref{eq:18}--\ref{eq:21}) clearly show that transitions between the states $|3\rangle$ and  $|2\rangle$ or between the states $|3\rangle$ and $|1\rangle$, are controlled by the driving field applied to the cavity field. However, the transition between the states $|2\rangle$ and $|1\rangle$ is dominated by the driving field applied to the qubit. Figure~\ref{fig5} numerically shows how the matrix elements change with the frequency $\omega_{d}$ and the strength $\Omega$ of the driving field. The nonzero values for $\text{Q}_{32}$, $\text{Q}_{31}$, and $\text{C}_{21}$, as shown in Fig.~\ref{fig4}(b, d, f), are limited by the first order of $g/\Delta$ which we omitted in Eqs.~(\ref{eq:7},\ref{eq:8}). Moreover, the behavior of $\text{Q}_{32}$, $\text{Q}_{31}$, and $\text{C}_{21}$ are identical to $\text{C}_{32}$, $\text{C}_{31}$, $\text{Q}_{21}$. So in the following analysis, we focus on the dominant matrix elements in the different parameter range.

(i) Outside the nesting regime, where $\omega_d<\omega_q-3\chi$. In this case,  the driving field applied to the qubit has no appreciable affects, and we have $|1\rangle\approx|g,0\rangle$, $|2\rangle\approx|e,0\rangle$, $|3\rangle\approx|g,1\rangle$, and $|4\rangle\approx|e,1\rangle$. As shown in Figs.~\ref{fig4}(a, c, e), $\text{C}_{32}\approx0$, $\text{C}_{31}\approx\text{Q}_{21}\approx1$.  The lowest three energy levels can be formed into a three-level system with $V$-type transitions. That is, one external field is applied to the cavity field to induce the transition between the states $|3\rangle$ and $|1\rangle$, while the other one is applied to the qubit to induce the transition between the states $|2\rangle$ and $|1\rangle$.

(ii) In the nesting regime, where $\omega_q-3\chi<\omega_d<\omega_q-\chi$,  the driving field applied to the qubit drastically changes the properties of the polariton states. We take $\text{C}_{32}$, shown in Fig.~\ref{fig5} (a), as an example. The saddle shape of $\text{C}_{32}$ is consistent with the boundary of the nesting regime. We first discuss the weak-driving case, i.e., $\Omega\approx0$. In this case, we have $|1\rangle\approx|g,0\rangle$, $|2\rangle\approx|e,0\rangle$, $|3\rangle\approx|g,1\rangle$, $|4\rangle\approx|e,1\rangle$, and this is the same with (i). The first sharp transition of $\text{C}_{32}$ occurs at $\omega_{|e,1\rangle}=\omega_{|g,0\rangle}$, when $\omega_d=\omega_q-3\chi$, and then the state $|3\rangle$ is changed to $|e,1\rangle$ with the change of the driving frequency $\omega_{d}$ through entering the nesting regime, the transition $\text{C}_{32}$, due to the driving field applied to the cavity, has a sudden jump from $0$ to the finite value. Then, when changing the frequency $\omega_{d}$ of the driving field,  when $\omega_d=\omega_q-\chi$, the state $|2\rangle$ changes from $|e,0\rangle$ to $|g,0\rangle$, while the state $|3\rangle$ can be approximated to $|3\rangle\approx|e,1\rangle$, hence the transition matrix element $\text{C}_{32}$ drops sharply. This is the same for the reverse trend of the transition matrix element $\text{C}_{31}$. It is understandable for the transition between the states $|2\rangle$ and $|1\rangle$, the sharp turning point is at the upper boundary, where $\omega_{|e,0\rangle}=\omega_{|g,0\rangle}$. In this condition, we have
$\text{C}_{32}\approx 1$, $\text{Q}_{21}\approx1$, and $\text{C}_{31}\approx0$ when the driving field is rather weak, i.e., $\Omega\approx0$, and the system behaves like a $\Xi$-type transitions where the state $|2\rangle $ and the state $|1\rangle$ is linked by the external field applied to the qubit, while the states $|3\rangle$ and $|2\rangle$ are linked by the external field applied to the cavity field.

(iii) In the nesting regime, when $\Omega\neq0$, with the increasing of the strength $\Omega$ of the driving field applied to the qubit, the state $|3\rangle$ becomes a mixing of the states $|g,1\rangle$ and $|e,1\rangle$, the state $|2\rangle$ becomes a mix of the states $|g,0\rangle$ and $|e,0\rangle$. Thus, the matrix element $\text{C}_{32}$ decreases gradually when increasing the driving strength $\Omega$. Likewise, the matrix element $\text{C}_{31}$ increases when increasing the driving strength $\Omega$. If two external driving field are applied to the cavity field, then two transitions between the states $|3\rangle$ and $|2\rangle$, and between the states $|3\rangle$ and $|1\rangle$ are induced, in this case we can construct a three-level system with the $\Lambda$-type transition, which will be used to study EIT and ATS in the following section. If another external field is applied to the qubit, then the transition between the states $|2\rangle$ and $|1\rangle$ can be induced, the three-level system now possesses cyclic transitions, or a $\Delta$-type~\cite{Liu2005a,Deppe2008} transition. For natural atoms, $\Delta$-type transitions do not exist, because the dipole operator possesses odd parity, it can only connect states with different parities. 

EIT only occurs in three-level systems with $\Lambda$-type transition or three-level systems with the upper driven $\Xi$-type transition~\cite{Lee2000,Abi-Salloum2010a}. In the next section, we will focus on a three-level system with $\Lambda$-type transition and study how EIT and ATS can be tuned by changing the driving field applied to the qubit.

%

\subsection{Tunable decay rates of mixed polarition states}

To study EIT, we first study how the classical driving field can be used to adjust the decay rates of the mixed polariton states by varying its amplitude and the frequency. The main idea is to change the ratio of how the cavity field or qubit contributes to the final mixture. If the cavity field and the qubit have different decay rates, then the decay rates of the polariton states vary with the weights of the cavity field branch and the qubit branch in the polariton states. To discuss the decay rates of the polariton states, let us assume that the environment interacting with the system can be described by bosonic operators. Then the Hamiltonian of the whole system, including the environment, can be written as
\begin{equation}
H^{\prime}=H_{S}+H_{E}+H_{I},\label{eq:22}
\end{equation}
 where the Hamiltonian $H_{S}$ is given in Eq.~(\ref{eq:1}). The free Hamiltonian $H_{E}$ in Eq.~(\ref{eq:22}) of the environment is given  by
\begin{eqnarray}
H_{E}&=&\hbar\int d\omega \omega b^{\dagger}(\omega)b(\omega)+\hbar\int d\omega^{\prime}\omega^{\prime} c^{\dagger}(\omega^{\prime})c(\omega^{\prime}).\label{eq:16-1}
\end{eqnarray}
The interaction Hamiltonian $H_{I}$ in Eq.~(\ref{eq:22}) between the system and the environment is given by
\begin{eqnarray}
H_{I}&=&\hbar\left[\int d\omega K\left(\omega\right) b^{\dagger}\left(\omega\right)a +\text{H.c.}\right] \nonumber \\
&+&\left[ \int d\omega^{\prime} \eta\left(\omega^{\prime}\right) c^{\dagger}(\omega^{\prime})\sigma_{-}+\text{H.c.}\right].\label{eq:17-1}
\end{eqnarray}
We have assumed that the environment of the cavity field is independent of that of the qubit. Here $b^{\dagger}(\omega)$ and $c^{\dagger}(\omega^{\prime})$ denote the creation operators of the environmental bosonic modes of the cavity field and the qubit, respectively. For simplicity, we further assume that the spectrum of the environment is flat, that is, both $K(\omega)$ and $\eta(\omega^{\prime})$ are independent of frequency. In this case, we can introduce the first Markov approximation
\begin{eqnarray}
K\left(\omega\right)&=&\sqrt{\gamma_{c}/2\pi},\\
\eta\left(\omega^{\prime}\right)&=& \sqrt{\gamma_{q}/2\pi}.
\end{eqnarray}

In the polariton basis,  the operators $a$ and $\sigma_{-}$ of the cavity field and the qubit can be expressed as
\begin{eqnarray}
a&=& \sum_{ij}\langle i|a|j\rangle \sigma_{ij},\\
\sigma_{-}&=& \sum_{lm}\langle l|\sigma_{-}|m\rangle \sigma_{lm},
\end{eqnarray}
where $|i\rangle$, $|j\rangle$, $|l\rangle$, and $|m\rangle$ denote mixed polariton states, which can be expressed by either the mixture of Eqs.~(\ref{eq:4}-\ref{eq:5}), for the general case, or the mixture of Eqs.~(\ref{eq:7}-\ref{eq:8}), for the large-detuning case. Here $\sigma_{ij}=|i\rangle\langle j|$. In the basis of the mixed polariton states, the interaction Hamiltonian $H_{I}$ can be rewritten as
\begin{eqnarray}
H_{I}&=&\hbar \int d\omega \sum_{ij}\left[\sqrt{\gamma_{ij}^{c}/2\pi} b^{\dagger}(\omega)\sigma_{ij}+\text{H.c.}\right] \nonumber\\
&+& \hbar\int d\omega^{\prime}  \left[\sum_{ij}\sqrt{\gamma^{q}_{ij}/2\pi}  c^{\dagger}(\omega^{\prime})\sigma_{ij}+\text{H.c.}\right],
\end{eqnarray}
with $\gamma_{ij}^{c}=\gamma_{c}|\langle i|a^\dagger|j\rangle|^2$ and $\gamma_{ij}^{q}=\gamma_{q}|\langle i|\sigma_{+}|j\rangle|^2$.  Thus, the total decay rate $\gamma _{ij}$ from one mixed polariton state $|i\rangle$ to another one $|j\rangle$ transition is given by
\begin{equation}\label{eq:30}
{\large\gamma _{ij}=\gamma_{ij}^{c}+\gamma_{ij}^{q}=\gamma _{c}\left|\langle{i}|a^{\dag }|{j}\rangle \right|^{2}+\gamma _{q}\left|\langle {i}|\sigma_{+}|{j}\rangle \right|^{2}}.
\end{equation}

In the large-detuning regime, where the cavity field and the qubit have very different frequencies, the decay rates, from one upper state to another lower state,  expressed from Eq.~(\ref{eq:11}) to Eq.~(\ref{eq:14}),  can be approximately given by
\begin{eqnarray}\label{eq:31}
\large \gamma_{31}& =& \gamma_{c}\sin^{2}\left(\frac{\theta_{u}+\theta_{l}}{2}\right),\\
\label{eq:32}
\large\gamma_{32}& =&  \gamma_{c}\cos^{2}\left(\frac{\theta_{u}+\theta_{l}}{2}\right),\\
\large \gamma_{21}& = &\gamma_{q}\cos^{4}\left(\frac{\theta_{l}}{2}\right).
\label{eq:33}
\end{eqnarray}
In the large-detuning regime, we also find $\gamma_{31}\approx \gamma_{42}$, $\gamma_{32}\approx \gamma_{41}$, $\gamma_{ii}^{c}=\gamma_{c}|\langle i|a^\dagger|i\rangle|^2\approx 0$, and $\gamma_{21}^{c}=\gamma^{q}_{43} \approx 0$.

\begin{figure}[hbt]
\includegraphics[scale=0.4]{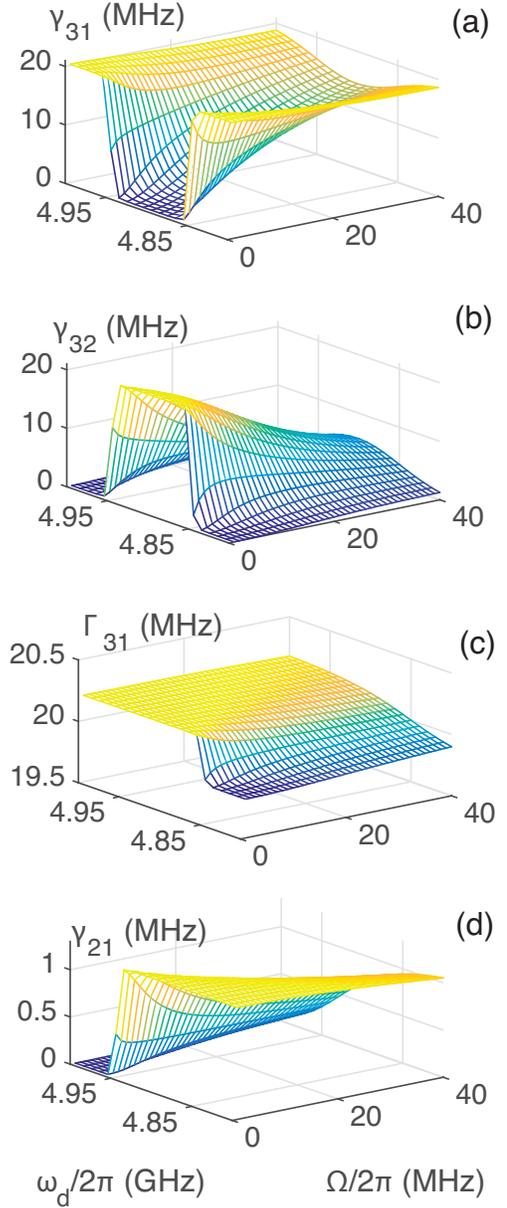}
\caption{ (color online). Decay rates $\gamma_{ij}$ versus both the frequency and strength of the driving field. Here, we set
$\Gamma_{31}=\gamma_{31}+\gamma_{32}\approx \gamma_c$. We have chosen $\gamma_{q}/2\pi$=1 MHz,
$\gamma_{c}/2\pi$=20 MHz. }
	\label{fig6}%
\end{figure}

Figure.~\ref{fig6} shows how the decay rates $\gamma_{ij}$ change with the frequency and strength of the driving field, plotted for the mixed polariton states. The decay rates are proportional to the square of the transition matrix elements. Therefore they have similar features for the dependence on the frequency and strength of the driving field. This can be seen by comparing Fig.~\ref{fig5} with Fig.~\ref{fig6}.
We define the total decay rate of the state $|3\rangle$ as $\Gamma_{31}=\gamma_{31}+\gamma_{32}$. We find that $\Gamma_{31}$ is hardly influenced by the driving field in the large-detuning case, except that there is a slight jump at the lower bound of the nesting regime. This is consistent with Eqs.~(\ref{eq:31}-\ref{eq:32}), i.e. $\Gamma_{31}\approx\gamma_c$.
In the nesting regime, the decay rates $\gamma_{31}$ and $\gamma_{32}$ change significantly when varying $\Omega$; the decay rate $\gamma_{21}$ is slightly decreased when $\Omega$ is increased. We find $\gamma_{31}=\gamma_{32}$ when $\Omega$ is taken as a particular value. This is an impedance-matching condition~\cite{Koshino2013a}, in which the two decay rates from the top energy level $|3\rangle$ to the two lowest energy levels $|1\rangle$ and $|2\rangle$ are the same, and microwave photons are down-converted efficiently
through Raman transitions due to impedance matching~\cite{Inomata2014}. Instead of using impedance-matching condition, below, we mainly study how EIT can occur in a chosen three-level system by using proper decay rates through adjusting the driving field.

\section{Electromagnetically induced transparency and Autler-Townes splitting}
\subsection{Linear response of a $\Lambda$ system}
We now study the EIT effect in a $\Lambda$ configuration atom interacting with two classical fields, as shown in Fig.~\ref{fig3}(b). The transition between the states $|3\rangle$ and $|2\rangle$ is linked by a strong external field with frequency $\omega_c$, hereafter called the control field. A probe field with frequency $\omega_p$ is applied to induce the transition between the states $|3\rangle$ and $|1\rangle$. The presence of a strong driving field dramatically modifies the response of the system to the weak probe field. As shown, for example, in Ref.~\cite{Fleischhauer2005}, the response of the probe field is analyzed using a semi-classical approach through the master equation.

The master equation for the reduced density matrix operator $\rho$ of the three-level system can be given by~\cite{Fleischhauer2005}
\begin{eqnarray}
\dot{\rho} &=&-\frac{i}{\hbar }[H_\text{int},\rho ]+\frac{\gamma _{31}}{2}[2\hat{%
\sigma}_{13}\rho \hat{\sigma}_{31}-\hat{\sigma}_{31}\hat{\sigma}_{13}\rho
-\rho \hat{\sigma}_{31}\hat{\sigma}_{13}] \nonumber\\
&+&\frac{\gamma _{32}}{2}[2\hat{\sigma}_{23}\rho \hat{\sigma}_{32}-\hat{%
\sigma}_{32}\hat{\sigma}_{23}\rho -\rho \hat{\sigma}_{32}\hat{\sigma}_{23}] \nonumber\\
&+&\frac{\gamma _{21}}{2}[2\hat{\sigma}_{12}\rho \hat{\sigma}_{21}-\hat{%
\sigma}_{21}\hat{\sigma}_{12}\rho -\rho \hat{\sigma}_{21}\hat{\sigma}_{12}]. \label{eq:34}
\end{eqnarray}
Here, we have neglected the energy-conserving dephasing processes of the three-level system. Also, $\gamma_{ij}$ is the spontaneous decay rate from $|i\rangle$ to $|j\rangle$, which coincides with Eq.~(\ref{eq:30}). Note that, compared to Ref.~\cite{Fleischhauer2005}, we have taken into account the spontaneous decay from $|2\rangle$ to $|1\rangle$. For natural atoms, $|2\rangle\rightarrow|1\rangle$ transition is forbidden, thus the dephasing rate of $|2\rangle$ dominates. However, in our compound system, we assume radiative decays of qubit and cavity are larger than dephasing processes~\cite{Koshino2013,Koshino2013a,Inomata2014}.  
$H_\text{int}$ describes the interaction of the three-level system with the control and probe fields in the interaction picture. In this system, $H_\text{int}$ can be given as
\begin{equation} \label{eq:35}
H_\text{int}=-\frac{\hbar}{2}\left(\Omega_{p}|3\rangle\langle1|e^{-i\Delta_1t}
+\Omega_{c}|3\rangle\langle2|e^{-i\Delta_2t}+\text{H.c.}\right),
\end{equation}
where $\Omega_{c}$ and $\Omega_{p}$ are the Rabi frequencies of the control and probe fields. We define the detunings as $\Delta_1=\omega_{31}-\omega_{p}$ and $\Delta_2=\omega_{32}-\omega_{c}$.
The master equation of the three-level system in Eq.~(\ref{eq:34}) can be solved using perturbation theory for the different orders of the strength of the probe field. We use the steady-state solution of the three-level system and assume that the three-level system is almost in the ground state, i.e.  $\rho_{11}\approx1$. Then, we find the linear susceptibility of the probe field $\chi ^{(1)}(-\omega _{p},\omega _{p})\propto \rho _{31}$. Omitting a multiplication factor, $\chi ^{(1)}(-\omega _{p},\omega _{p})$ can be given as~\cite{Abi-Salloum2010a,Anisimov2011}
\begin{equation}\label{eq:36}
\chi ^{\left(1\right)}\left(-\omega _{p},\omega _{p}\right)=\frac{\delta -\frac{i\gamma _{21}}{2}}{\left(\delta  -\frac{i\Gamma
		_{31}}{2}\right)\left(\delta+\Delta_2 -\frac{i\gamma _{21}}{2}\right)-\frac{\Omega _{c}^{2}}{4}}.
\end{equation}
Here $\delta=\Delta_1-\Delta_2$ is the two-photon detuning. The total decay rate $\Gamma_{31}$ of the state $|3\rangle$ is defined as
\begin{equation}
\Gamma_{31}=\gamma_{31}+\gamma_{32}.
\end{equation}
Equation~(\ref{eq:36}) is the starting point for the discussions on the difference between EIT and ATS as in Refs.~\cite{Abi-Salloum2010a,Anisimov2011,SunHuiChen2014,Peng2014}. 

Note that in Eq.~\ref{eq:36}, if we take into account the dephasing processes of states $|3\rangle$ and $|2\rangle$ with rates $\gamma_\text{3deph}$ and $\gamma_\text{2deph}$, and neglect the spontaneous decay from $|2\rangle$ to $|1\rangle$ in the master equation, then the coherence decay rates are  $\Gamma_{31}=\gamma_{31}+\gamma_{32}+\gamma_\text{3deph}$, $\Gamma_{32}=\gamma_{31}+\gamma_{32}+\gamma_\text{3deph}+\gamma_\text{2deph}$, $\gamma_{21}=\gamma_\text{2deph}$. These are the definitions used in Refs.~\cite{Murali2004,Fleischhauer2005,Anisimov2011}.

\subsection{Difference between EIT and ATS}
To shed light on the difference between EIT and ATS, we follow the spectral decomposition method as used in Refs.~\cite{Agarwal1997,Anisimov2008,Anisimov2011,Fleischhauer2005,Peng2014}. For simplicity, we assume that $\omega_c$ is resonant with $\omega_{32}$, i.e., $\Delta_2=0$. The imaginary part of the linear susceptibility $\chi$ characterizes the absorption, which can be decomposed into two resonances,
\begin{equation}
{\rm Im}\left(\chi\right)={\rm Im}\left(\frac{\chi_{+}}{\delta-\delta_{+}}+\frac{\chi_{-}}{\delta-\delta_{-}}\right)
\end{equation}
where $\chi_\pm=\pm{\left(\delta_\pm-{i\gamma _{21}}/{2}\right)}/{\left(\delta_+-\delta_-\right)}$. The poles of the denominator are given by
\begin{equation}\label{eq:39}
\delta_{\pm}=i\frac{\Gamma_{31}+\gamma_{21}}{4}\pm\frac{1}{2}\sqrt{\Omega_{c}^{2}-\frac{1}{4}\left(\Gamma_{31}-\gamma_{21}\right)^{2}}.
\end{equation}
Equation~(\ref{eq:39}) gives the threshold value for EIT:
\begin{equation}\label{eq:40}
|\Omega_{c}|=\frac{1}{2}|\Gamma_{31}-\gamma_{21}|.
\end{equation}

(i) When $|\Omega_{c}|\gg|\Gamma_{31}-\gamma_{21}|/2$, i.e., the strong-controlling field case, ATS occurs. The final spectrum of ${\rm Im}(\chi)$ is decomposed of two positive Lorentzians with equal linewidths. They are separated by a distance proportional to $\Omega_c$~\cite{Abi-Salloum2010a,Anisimov2011}.

(ii) When $|\Omega_{c}|<|\Gamma_{31}-\gamma_{21}|/2$, EIT occurs. The major characteristics of EIT is that the absorption spectrum is composed of one broad positive Lorentzian and one narrow negative Lorentzian. Both are centered at $\delta=0$. They cancel each other and result in the reduction of absorption to the probe field~\cite{Abi-Salloum2010a,Anisimov2011}.

For the ideal three-level system with $\Lambda$-type transitions, the transition between $|2\rangle$ and $|1\rangle$ is forbidden, thus $\gamma_{21}=0$, which is easy to find in natural atomic systems. However, $\gamma_{21}$ is usually nonzero in artificial atomic systems. Thus, to observe an absorption dip with a nonzero value of $\gamma_{21}$, we must require $\gamma_{21}\ll\Gamma_{31}$, otherwise the dip in the absorption spectrum is absent~\cite{Fleischhauer2005}.

\subsection{EIT and ATS in polariton system}
We now turn to study how the EIT and ATS can be realized in polariton systems by adjusting the driving field. As shown above, in the nesting regime, when $\Omega\neq0$, the polariton system can be used to construct an effective three-level system with $\Lambda$-type transitions, where the state $|3\rangle$ can be linked to $|2\rangle$ and $|1\rangle$ by applying the external fields to the cavity mode. Here, we assume that a strong controlling field $\text{A} _{c}^{\prime}\cos(\omega_{c}t)$ and a weak probe field $\text{A}_{p}^{\prime}\cos(\omega_{p}t)$ are applied to the system through the cavity mode. Where $\omega_{c}$ ($\omega_{p}$) and $A_{c}^{\prime}$ ($A_{p}^{\prime}$) are the frequency and amplitude of the controlling (probing) field. Under the rotating wave approximation, the Hamiltonian between the cavity mode and the two external fields can be written as
\begin{equation}
H_\text{drive}=-\frac{\hbar}{2}\left(\text{A}_{p}a^{\dagger}e^{-i\omega_{p}t}
+\text{A}_{c}a^{\dagger}e^{-i\omega_{c}t}+\text{H.c.}\right).
\end{equation}
Here, the coupling strength $A_{c}$ ($A_{p}$) between the controlling (probing) field and the cavity field is proportional to the amplitude $A_{c}^{\prime}$ ($A_{p}^{\prime}$) of the controlling (probing) field.

Similar to the three-level systems for demonstrating EIT and ATS, we now assume that the controlling field is used to induce the transition between the states $|3\rangle$ and $|2\rangle$, while the probe field is used to induce the transition between the states $|3\rangle$ and $|1\rangle$,
then in the mixed polariton state basis, using Eqs.~(\ref{eq:18}--\ref{eq:19}), the expressions for the Rabi frequencies $\Omega_{c}$ and $\Omega_{p}$, shown in Eq.~(\ref{eq:35}), become
\begin{eqnarray}
|\Omega_{c}|&\approx&{\text{A}_{c}}\text{C}_{32}, \label{eq:42}\\
|\Omega_{p}|&\approx&{\text{A}_{p}}\text{C}_{31}.
\end{eqnarray}
Figure~\ref{fig5} shows that the transition matrix element $\text{C}_{32}$ decreases while $\text{C}_{31}$ increases
in the nesting regime, when the strength $\Omega$ of the driving field is increased. Both of them are in the range of 0 to 1.

Now we turn to study the threshold of EIT set by Eq~(\ref{eq:40}) in the polariton system. With the help of Eqs.~(\ref{eq:31}-\ref{eq:33}), we obtain,
\begin{eqnarray}
\Gamma _{31}&=&\gamma _{31}+\gamma _{32}=\gamma _{c},\\
\gamma _{21}&=&\large\gamma _{q}\cos ^{4}\left(\frac{\theta _{l}}{2}\right).
\label{eq:41}
\end{eqnarray}
It is clear that the driving field can hardly affect $\Gamma_{31}$, as shown in Fig~\ref{fig6} (c). For EIT in a $\Lambda$ system~\cite{Fleischhauer2005}, $\gamma _{21}$ should be negligible, and this requires $\gamma_c\gg\gamma_q$. Therefore, in order to achieve EIT in our polariton system, the Rabi frequency of the controlling field should satisfy the condition
\begin{equation}
|\Omega_c|<\frac{\gamma_c}{2}.\label{eq:46}
\end{equation}

To investigate EIT and ATS in our polariton system, we now choose $\omega_{d}=4.9$ GHz, $\gamma_{c}=20$ MHz, $\gamma_{q}=1$ MHz in the following. In Table~\ref{table1}, we show explicitly how the transition matrix elements and energy level spacing change with $\Omega$ in the nesting regime.
\begin{table}[h]
    \setlength{\tabcolsep}{5pt}
	\begin{tabular}{|c|c|c|c|c|c|c|c|c|c|}
		\hline  $\Omega$ & $C_{31}$ & $C_{32}$ & $Q_{21}$ & $Q_{31}$ & $Q_{32}$ & $C_{21}$ & $\omega_{21}$ & $\omega_{32}$  & Type \\
		\hline
		0 & 0 & 1 & 1 & 0 & 0.1 & 0.1 & 54 & 5050 & $\Xi$ \\
		\hline  10 & 0.37 & 0.93 & 0.96 & 0 & 0.1 & 0.1 & 59 & 5047 &  $\Lambda$,$\Delta$ \\
		\hline 20 & 0.62 & 0.77 & 0.89 & 0 & 0.1 & 0.09 & 66 & 5037 &  $\Lambda$,$\Delta$ \\
		\hline 30 & 0.77 & 0.64 & 0.82 & 0 & 0.1 & 0.08 & 78 & 5023 &  $\Lambda$,$\Delta$ \\
		\hline 40 & 0.85 & 0.53 & 0.76 & 0 & 0.1 & 0.08 & 89 & 5007 &  $\Lambda$,$\Delta$ \\
		\hline
	\end{tabular}
	\caption{Numerical calculations for the matrix elements and transition frequencies. Here we choose $\omega_d=4.9$ GHz in the middle of the nesting regime. The units of $\Omega$, $\omega_{21}$, and $\omega_{32}$ are in MHz.  Other parameters are the same as in Fig.~\ref{fig4}.}\label{table1}
\end{table}

\begin{figure}[hbt]
	\includegraphics[scale=0.32]{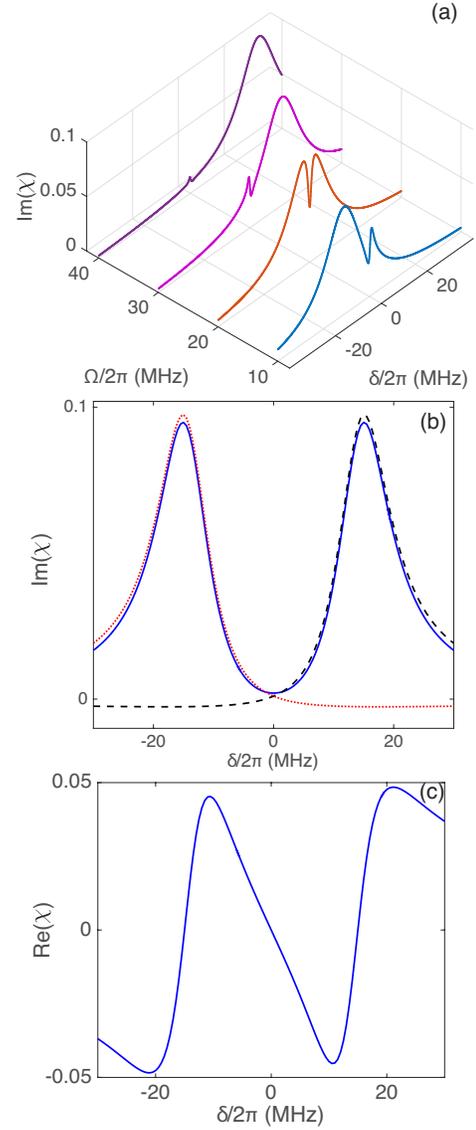}
	\caption{ (color online). (a) The imaginary part of the susceptibility, Im$(\chi)$, versus the strength of the driving field $\Omega$ and the two-photon detuning $\delta$, when the ATS condition is satisfied. Here, we have
		chosen $\omega_d=4.9$ GHz, $\gamma_c=20$ MHz, $\gamma_q=1$ MHz, and $\text{A}_c=30$ MHz, while the rest of the parameters are identical to the ones in Fig~\ref{fig4}. The control field frequency is $\omega_c=5.037$ GHz, which is resonant with $\omega_{32}$ for the $\Omega=20$ MHz case. (b) shows the spectral decomposition of ${\rm Im}(\chi)$ at resonance, i.e. $\Delta_2=0$. Here, the blue solid curve corresponds to the absorption spectrum. The red-dotted and the black-dashed curves correspond to two Lorentzian profiles. (c) shows the real part of $\chi$  characterizing the refractive properties at resonance. }\label{fig7}
\end{figure}

\begin{figure}[hbt]
\includegraphics[scale=0.32]{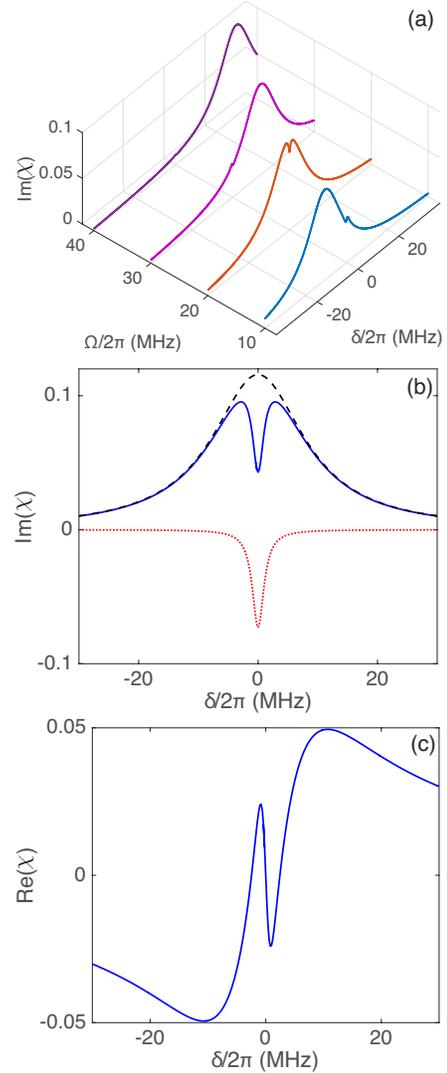}
\caption{ (color online). (a) The imaginary part of the susceptibility, Im$(\chi)$, versus the strength of the driving field $\Omega$ and the two-photon detuning $\delta$, when the EIT condition is satisfied. Here, we have chosen $\omega_d=4.9$ GHz, $\gamma_c=20$ MHz,$\gamma_q=1$ MHz, and $\text{A}_c=5$ MHz, while the rest of the parameters are identical to the ones in Fig~\ref{fig4}. The control field frequency is $\omega_c=5.037$ GHz, which is resonant with $\omega_{32}$ for $\Omega=20$ MHz case. (b) shows the spectral decomposition of ${\rm Im}(\chi)$ at resonance, i.e., $\Delta_2=0$. Here, the blue solid curve corresponds to the absorption spectrum. The red-dotted and the black-dashed curves correspond to two Lorentzian profiles. (c) shows the real part of $\chi$ at resonance, which characterizes the refractive properties. }
\label{fig8}
\end{figure}

In Fig.~\ref{fig7} and Fig.~\ref{fig8}, we set the frequency $\omega_c=5.037$ GHz of the controlling field, which is resonant with $\omega_{32}$ when $\Omega=20$ MHz, as shown in Table~\ref{table1}. From Eqs.~(\ref{eq:42}) and (\ref{eq:46}) and Table~\ref{table1}, we can find that the polariton system satisfies the ATS condition (EIT condition) when  $\text{A}_c=30$ MHz ($\text{A}_c=5$ MHz) for $\gamma_{c}=20$ MHz, $\gamma_{q}=1$ MHz, with the value of $\Omega$ in the range $10$ MHz to $40$ MHz. However, we find that ATS and EIT also depend on the strength $\Omega$ of the driving field. We plot $\Omega=10,20,30,40$ MHz cases separately in Fig.~\ref{fig7} and ~\ref{fig8}, in which the parameters are consistent with Table~\ref{table1}.

When the ATS condition is satisfied for $\text{A}_c=30$ MHz, Figure~\ref{fig7}(a) shows how the absorption spectra through ${\rm Im}(\chi)$ varies with the strength $\Omega$ of the driving field.  Figure~\ref{fig7}(b) shows the variations of spectral decomposition of ${\rm Im}(\chi)$ with two photon detuning $\delta$ at resonance with two positive Lorentzian shape spectra.  Figure~\ref{fig7}(c) shows the variations of the real part, ${\rm Re}(\chi)$, of the susceptibility $\chi$ with $\delta$. Clearly, when varying $\Omega$, the absorption spectra can have two symmetric or asymmetric peaks. The asymmetries are mainly caused by the non-zero detuning between $\omega_c$ and $\omega_{32}$, i.e., $\Delta_2\neq0$ in Eq.~(\ref{eq:36}). Because we have assumed that the frequency $\omega_{c}=5.037$ GHz of the controlling field, which is resonant with $\omega_{32}$ only when $\Omega=20$ MHz. In other values of $\Omega$, the controlling field is nonresonant with $\omega_{32}$, which decreases when $\Omega$ is increased, as shown in Fig.~\ref{fig4}(b) and Table~\ref{table1}. Therefore, the windows of two peaks disappear for a given $\omega_{c}$ when $\Omega$ becomes very large. We emphasize that the windows with two peaks can always be found for a given $\Omega$ by varying the frequency $\omega_{c}$.

Figure~\ref{fig8} shows how the imaginary and real parts of the susceptibility $\chi$ vary with the strength $\Omega$ and the two photon detuning $\delta$ when the EIT condition is satisfied for $\text{A}_c=5$ MHz.  The two asymmetric peaks in the spectrum in Fig.~\ref{fig8}(a) are also due to the non-zero detuning between $\omega_c$ and $\omega_{32}$.  Figure~\ref{fig8}(a) also shows that the transparency windows not only depend on the strength $A_{c}$ of the control field but also depend on the strength $\Omega$ of the driving field. When the strength $\Omega$ of the driving field becomes very strong, the transparency windows disappear even the EIT condition $\Omega_{c}<\gamma_{c}/2$ is satisfied for a given $\omega_{c}$. Same as ATS, we can always find transparency windows by changing $\omega_{c}$ for a given $\Omega$ when the EIT condition is satisfied.
Figure~\ref{fig8}(b) clearly shows that the reduction of absorption is caused by the cancellation of positive and negative
Lorentzian profiles. Compared with ATS in Fig.~\ref{fig7}(b), the transmission window is sharper and the width is less than $\Gamma_{31}$, this is due to interference effects~\cite{Fleischhauer2005}. The real part, ${\rm Re}(\chi)$, characterizing refractive properties shown in Fig.~\ref{fig8} (c), varies much more rapidly in the transparency window in contrast with that in Fig.~\ref{fig7}(c).

Note that Ref.~\cite{Anisimov2011} analyzes the threshold for EIT and ATS only for the case for $\Delta_2=0$. If $\Delta_2$ becomes large, the Raman model has to be taken into account. In Raman model the spectral decomposition becomes one broad Lorentzian at the center $\delta=0$ with another narrow Lorentzian at $\delta=\Delta_2$~\cite{Anisimov}. This is different from both EIT and ATS.

\section{Discussions and Conclusions}

We studied how to achieve EIT and ATS in a driven superconducting circuit QED system. Without the driving field, the system is reduced to the Jaynes-Cummings model. EIT based on the dressed states of the Jaynes-Cummings model was studied in Ref.~\cite{Ian2010}. In contrast to Ref.~\cite{Ian2010}, where the decay rates cannot be changed once the sample is fabricated, we introduce an additional driving field to form a three-level system to study EIT and ATS. That is, the three-level system for EIT and ATS is formed by polaritons, which is the doubly-dressed qubit states through a cavity field and a classical driving field. It is known that the polaritons are hybridization of the states of both the qubit and the cavity field, thus their decay rates include both contributions of the cavity field and the qubit. The qubit and the single-mode cavity field have independent decay rates, and also the weights of the cavity field state and the qubit state in the polaritons can be adjusted by the driving field. Thus, the decay rates of the chosen three-level system can be adjusted by the driving field. Therefore, it is easy to find a parameter regime to realize EIT, and also demonstrate the transition from EIT to ATS.

In particular, we have provided a detailed study of how EIT and ATS can be demonstrated in a so-called nesting regime~\cite{Koshino2013a} by varying the driving field, when the qubit and the cavity field are in the large detuning regime. We find that the driving field can also be used to control windows between the two peaks of EIT or ATS. Sometimes, we can only find a peak and cannot find a windows even when then EIT and ATS conditions are satisfied for a given frequency of the control field. This is because the driving energy structure of the chosen three-level system are changed by the driving field, when the frequency of the control field is largely out of resonance with the two addressed energy levels, the two peaks become one peak and then the transparency window disappears. To observe a dip in the absorption spectrum for both EIT and ATS, it is also required that the qubit decay rate is negligibly small compared with the cavity decay rate.

Finally, we would like to mention that our proposed three-level system can also possess $\Delta$-type, $\Xi$-type, and $V$-type transitions by using different configurations of the external fields applying to the driven circuit QED system. Thus, this system provides a very good platform to demonstrate various atomic and quantum optical phenomena. For a single artificial atom, the decay rates are intrinsic properties and are very hard to control. However, our compound system can be manufactured by tailoring the qubit and cavity decay. This can be helpful in guiding future experimental observation of EIT in driven circuit QED systems. The parameters for the numerical calculations are taken from accessible experimental data; thus our proposal should be experimentally realizable with current superconducting quantum circuits.

Note added: after this work was completed, an experiment observed EIT in a SQC system~\cite{Novikov2015}.
\begin{acknowledgments}
We thank P. M. Anisimov, A. F. Kockum, and A. Miranowicz for useful comments on the manuscript.
Y.X.L. acknowledges the support of the National Basic Research Program of
China Grant No. 2014CB921401 and the National Natural Science Foundation of
China under Grant No. 91321208. F.N. is partially supported by the
RIKEN iTHES Project,
the MURI Center for Dynamic Magneto-Optics
via the AFOSR award number FA9550-14-1-0040,
the IMPACT program of JST,
and a Grant-in-Aid for Scientific Research (A).

\end{acknowledgments}


\end{document}